\documentclass[preprint,12pt]{elsarticle}

\usepackage{amssymb}

\usepackage[latin1]{inputenc}
\usepackage{amsmath}
\usepackage{amsfonts}
\usepackage{amssymb}
\usepackage{graphicx}
\usepackage{subfig}
\usepackage{mathrsfs}
\usepackage{listings}
\usepackage{verbatim}
\usepackage{mathtools}
\usepackage{simplewick}
\usepackage{color}
\usepackage[left=1.5cm,right=1.5cm,top=2cm,bottom=2cm]{geometry}\usepackage{subfig}

\newcommand{\be}{\begin{equation}}
\newcommand{\ee}{\end{equation}}
\newcommand{\bq}{\begin{eqnarray}}
\newcommand{\eq}{\end{eqnarray}}

\newcommand{\one}{\hbox{\rm 1\kern-.27em I}}

\journal{Physics Letters B}

\begin{document}

\begin{frontmatter}


\author{James P. Edwards and Paul Mansfield \\Centre for Particle
Theory, University of
Durham, Durham DH1 3LE, UK \\ Email:
J.P.Edwards@durham.ac.uk,
P.R.W.Mansfield@durham.ac.uk}

\title{QED as the tensionless limit of the spinning string with contact interaction        DCPT-14/39}


\author{}

\address{}

\begin{abstract}
We outline how QED with spinor matter can be described by the tensionless limit of spinning strings with contact interactions. The strings represent electric lines of force with charges at their ends. The contact interaction is constructed from a delta-function on the world-sheet which, although off-shell, decouples from the world-sheet metric. Integrating out the string degrees of freedom with fixed boundary generates the super-Wilson loop that couples spinor matter to electromagnetism in the world-line formalism. World-sheet and world-line, but not spacetime, supersymmetry underpin the model.
\end{abstract}


\end{frontmatter}


\section{Introduction}
Quantum Electrodynamics is perhaps the most successful physical theory to confront experiment, and so it might seem redundant to consider an alternative formulation. However, as an Abelian gauge theory it is a simpler version of the non-Abelian gauge theory of the Standard Model to which new approaches may still be of interest. In this letter we treat QED by taking the electric lines of force as the basic degrees of freedom of the electromagnetic field. This immediately requires the technology of string theory but applied to a non-standard setting in which the ends of the lines of force are electrically charged particles and the electromagnetic interaction becomes a contact interaction described by $\delta$-functions on the world-sheet. We will show that even though these are off-shell they can be constructed to be independent of the scale of the world-sheet metric because of the non-standard boundary conditions. Unwanted divergences that might occur when there is more than one interaction on each world-
sheet are eliminated when the model has world-sheet supersymmetry and this allows the interaction to be exponentiated thus generating the super-Wilson loops that couple spinor matter to the electromagnetic field on the world-sheet boundaries. Including supersymmetric boundary terms in the action quantises the spinor matter in the world-line formalism. QED emerges in the tensionless limit, so that the strings representing the lines of force are potentially large, although string corrections might set in at very large length scales.

Conventionally, the first step in the passage to the quantum theory from the classical Maxwell equations
\be
\epsilon^{\mu\nu\lambda\rho}\,\partial_\nu  F_{\lambda\rho}=0,\quad
\partial^\mu \, F_{\mu\nu}=J_\nu \,,
\ee
is to solve the first set by introducing a gauge potential, $A$,  and then construct a Lagrangian with this as the dynamical variable (modulo gauge transformations) so that the second set appear as Euler-Lagrange equations. We choose the alternative starting point by solving the second set. For simplicity we consider a system consisting 
of particle anti-particle pairs created and then mutually annihilating, so the current density is 
\be
J^\mu(x)=\sum q\int_{B}\delta^4( x-w)\,dw^\mu
\label{curr}
\ee
where the world-lines $B$ are closed.
{\it One} solution is to take
\be
F_{\mu\nu}(x)=\sum
-q\int_\Sigma \delta^4(x-X)\, d\Sigma_{\mu\nu}(X)\,,
\label{Fsolnn}
\ee
where $d\Sigma_{\mu\nu}$ is an element of area on a surface $\Sigma$ spanning $B$. This field-strength, which vanishes away from $\Sigma$, may be interpreted as that of a single line of force. We will take this surface $\Sigma$ as the dynamical degree of freedom instead of the gauge potential. Treating this as the basic physical object is reminiscent of Faraday's approach to electromagnetism\cite{Faraday} in which lines of force are the fundamental degrees of freedom. This was echoed in Dirac's 1955 proposal\cite{Dirac} that creation operators for electric charges should simultaneously create part of the electromagnetic field so that the radially symmetric Coulomb field for a single charge would emerge from quantum mechanical averaging of (\ref{Fsolnn}). An equivalent expression to  (\ref{Fsolnn}) was used to describe the polarisation vector of charged matter for molecular electrodynamics \cite{Woolley} and in the context of non-linear electrodynamics by Nielsen and Olsen\cite{NO} to form a field theory 
describing the dual string. Its dual is also present in theories of electromagnetism with magnetic monopoles \cite{diracM} and  has been used \cite{Nambu, Baker} to derive an effective string theory describing the evolution of the Dirac string linking two such poles.

Substituting into the classical electromagnetic action gives
\bq
S_{EM}&=&-{1\over 4}\int d^4x~ F_{\mu\nu}F^{\mu\nu}\nonumber\\
&=&{q^2\over 4}\delta^2(0) \,{\rm Area}(\Sigma)+{q^2\over 4}\int_\Sigma\left.
d\Sigma^{\mu\nu}(\xi)\,\delta^4\left(X(\xi)-X(\xi')\right)\,
d\Sigma_{\mu\nu}(\xi')\right|_{\xi\neq\xi'}
\label{Sem}\nonumber\\
\eq
the first term is proportional to the Nambu-Goto action, albeit with a divergent coefficient, whilst the second is a
self-intersection interaction. Clearly, to proceed further requires the machinery of string theory but with non-standard contact interactions rather than conventional splitting and joining. Similar interactions have previously been discussed by Kalb and Ramond \cite{kalbR} and the one we propose here satisfies the consistency constraints they derive. This action has been applied classically \cite{NRH} to the problem of confinement but without self-intersections or quantisation.

In \cite{Mansfield:2011eq} it was shown that the average of (\ref{Fsolnn}) over $\Sigma$ constructed according to Polyakov's approach to the bosonic string\cite{Polyakov} does in fact yield the electromagnetic field generated by $J^\mu$, Wick rotated to Euclidean signature where the functional integrals behave better:
\be
4\pi^2\langle \int_\Sigma \delta^4(x-X)\,
d\Sigma_{\mu\nu}(X)\rangle _{\Sigma}
=\partial_\mu \int_B {dw_\nu \over ||x-w||^2}-\partial_\nu
\int_B {dw_\mu \over ||x-w||^2}
\label{sol}
\ee
where the average over $\Sigma$ of any functional
$\Omega[\Sigma]$ is
\be
\langle \Omega\rangle_\Sigma={1\over Z}\int {\mathscr{D}}g\,{\mathscr{D}_{g}}{X}\,\Omega\,e^{-S[X,\, g]}\,,\quad
S[X,g]={
{1\over 4\pi\alpha'}\int_D g^{ab}\,{\partial{
X^\mu}\over\partial\xi^a}{\partial{
X^\mu}\over\partial\xi^b}\,\sqrt {g}\,d^2\xi}
\,.\label{avvS'}
\ee
Remarkably this result is independent of the scale of the world-sheet metric despite the $\delta$-function being off-shell and is also independent of the string tension, $\alpha'$. Integrating over a different surface $\Sigma'$ spanning the fixed closed loop $B'$ gives
\be
\langle \int_{\Sigma \Sigma'}d\Sigma^{'\mu\nu}
\delta^4(x-X)\,
d\Sigma^{\mu\nu}(X)\rangle _{\Sigma}
= {1\over 2\pi^2}\int_{BB'} {dw'\cdot dw \over ||w'-w||^2}
\label{sol'}
\ee
(since the right-hand-side is independent of this second surface we could obtain a more symmetrical looking result by also averaging over $\Sigma'$). The right-hand-side is the electromagnetic interaction between the two loops of charges $B$ and $B'$. If it were possible to show that this exponentiates
then we would be able to express the expectation value of Wilson loops in Maxwell theory, i.e. 
\be
\int {{\mathscr{D}}A\over N}\,e^{-S_{gf}}\,\prod_j
e^{-iq\oint_{B_j}dw\cdot A}\label{expWL}
\ee
(where $S_{gf} $ is the usual gauge-fixed action for the electromagnetic field)
 as the partition function of  first quantised strings with fixed boundaries and which interact on contact:
\be
\int\left( \prod_{j} {\mathscr{D}(X_j,g_j)\over Z_0}\right)\, e^{-S},
\ee
where
\be
S=\sum_j S[X_j,g_j]+\sum_{jk} q^2 \int_{\Sigma_j \Sigma_k}d\Sigma^{\mu\nu}_j\,
\delta^4(X_j-X_k)\,
d\Sigma_{\mu\nu}^k\label{actb}
\ee
effectively replacing the quantised electromagnetic field by quantised strings with fixed boundaries. Integrating over the boundaries with appropriate weights quantises the charged sources along the lines of Strassler's world-line approach\cite{Strassler:1992zr} (see also \cite{Schubert:2001he} and \cite{Ilderton:2014mla} for recent applications)
so we would arrive at a reformulation of QED in terms of strings with unusual boundary terms and contact interactions. This programme is pursued in detail in \cite{us} where it is shown that with bosonic matter the programme is somewhat challenging, but that for spinor matter the additional structure resulting from a spinning world-sheet renders the approach tractable, and it is this, actually more realistic, case that we describe in this letter.

Evaluating the conventional QED functional integral by first integrating over spinor matter results in the fermionic determinant depending on the gauge field $A_\mu$. Strassler represents this determinant by a world-line functional integral. We will use a reparametrisation invariant formulation based on the action of Brink, di Vecchia and Howe\cite{Brink:1976uf} (further details are given in \cite{us})
\be
{\rm ln\,Det}\left(-\left(\gamma\cdot(\partial+iA)\right)^2+m^2\right)=-
\int {\mathscr{D}} (h,w,\chi,\psi)\,W_s[A]\,e^{-S_{BdVH}}\label{wlf}
\ee
where
\be
S_{BdVH}=
\oint\left({1\over 2\sqrt{h}}\,\left({d{ w}\over 
dx}\right)^2-i\psi\cdot {d \psi\over dx}-
i{\chi \over\sqrt h} {d w\over dx}\cdot \psi\right)\,\,dx 
\ee
(for simplicity we drop the mass term)
and $W$ is the supersymmetric Wilson loop
\be
W_s[A]={\rm exp}\left(i\oint \left ({dw\over dx} \cdot A+{1\over 2}F_{\mu\nu}\psi^\mu\psi^\nu\sqrt h \right)\,dx\right)\,.
\ee
Here $\chi$ is the fermionic partner of  $h$ which is an intrinsic metric on the world-line parametrised by $\xi$, and $\psi^\mu$ are fermionic partners of the co-ordinates, $w^\mu$ in d-dimensional space-time.
$w$, $h^{1/4}$ and $\chi$ have dimensions of length but $\psi$ is dimensionless, so  $S_{BdVH}$ is dimensionless as well.
 As is well-known, the action $S_{BdVH}$ and the exponent of $W$ have the worldline supersymmetry 
\be
\delta w=i\alpha\psi\,,\quad\delta\psi={\alpha\over\sqrt h}\left({dw\over dx}-{i\over 2}\chi\psi\right)\,,\quad
\delta\sqrt h=i\alpha\chi\,,\quad \delta\chi= 2{d\alpha\over dx}\,,\label{wlsusy}
\ee
despite the absence of supersymmetry in the spacetime theory of QED. Curiously the fermionic Green function may also be expressed in the same form of the right-hand-side of (\ref{wlf}) but using open worldlines\cite{us} with appropriate conditions at their ends. In (\ref{wlf}) the gauge-field, $A$, appears only in $W$ so to complete the quantisation of QED it just remains to functionally integrate over $A$ using the super-Wilson loop equivalent of (\ref{expWL}). It is our purpose to show that this last step can be replaced by a functional integral over spinning strings spanning the closed loops $B$, so that together with the boundary action $S_{BdVH}$ and contact interactions we arrive at a string theory reformulation of QED.
\section{The interacting string theory}
The spinning string has gauge-fixed action
\be
S_{spin}=
{1\over 4\pi\alpha'}\left(\int_H d^2z\,d^2\theta ~\bar D {\bf X}\cdot D{\bf X}-\int_{y=0} dx\,\bar\Psi\cdot\Psi\right)
\ee
where we take the parameter domain to be the upper-half complex $z=x+iy$-plane. $\theta$ and $\bar\theta$ are anti-commuting variables that enter the derivative operators $D=\partial/\partial\theta +\theta\partial/\partial z$ and $\bar D=\partial/\partial\bar\theta +\bar\theta\partial/\partial \bar z$ with $\partial/\partial z=(\partial/\partial x-i\partial/\partial y)/2$. $d^2z=-2i\, dx \,dy$ and Stokes' theorem
becomes $\int d^2z\,d^2\theta\, DF=\oint d\bar z d^2\theta\, \theta F$ and $\int d^2z\,d^2\theta \,\bar DF=-\oint dz d^2\theta\, \bar\theta F$. Since we work exclusively with functional integrals we assume a Wick rotation to Euclidean spacetime of dimension $d$.  The superfield has components
\be
{\bf X}=X+\theta\Psi+\bar\theta\bar\Psi+\bar\theta\theta B
\ee
with B an auxiliary field. $X$, $\Psi$, $\bar\Psi$ and $\sqrt {\alpha'}$ have dimensions of length.
We impose Dirichlet boundary conditions that relate $\bf X$ on $y=0$ to the world-line variables
\be
\left.X\right|_{y=0}=w,\quad \left.\left(\Psi+\bar\Psi\right)\right|_{y=0}=h^{1/4}\,\psi\,.\label{bcon}
\ee
The factor of $h^{1/4}$ is necessary since $\psi$ is a world-line scalar.  
The first term in the action is standard\cite{Friedan:1985ge}. We have added a boundary term (that would vanish under the usual Neveu-Schwarz or Ramond boundary conditions) to ensure invariance under the residual global supersymmetry
\be
\delta{\bf X}=\eta\left({\partial\over\partial\theta}-\theta{\partial\over\partial z}
+{\partial\over\partial\bar\theta}-\bar\theta{\partial\over\partial \bar z}
\right){\bf X}\label{susy}
\ee
which also acts on the world-line variables (with $\alpha=h^{1/4}\eta$ in (\ref{wlsusy})) to preserve the boundary conditions and $S_{BdVH}$. 

Consider now a number of spinning strings, each spannning a closed boundary and interacting on contact with each other with an action that is the generalisation of (\ref{actb})
\bq
&&S_s=\sum_j S_{spin}[{\bf X}_j]+\sum_{jk}S_{int}[{\bf X}_j,{\bf X}_{k}]
\label{sact}\\
&&S_{int}=q^2\int d^2z_jd^2\theta_{j}\left(\bar D {\bf X}^{[\mu} D{\bf X}^{\nu]}-\delta(y)\theta\bar{\theta}{\bar\Psi}^{[\mu} {\Psi}^{\nu]}\right)_j 
\delta^d({\bf X}_j-{\bf X}_{k})\,d^2z_{k}d^2\theta_{k}\left( \bar D {\bf X}^{[\mu} D{\bf X}^{\nu]}-\delta(y)\theta\bar{\theta}{\bar\Psi}^{[\mu} {\Psi}^{\nu]}\right)_{k}\nonumber
\eq
This too is invariant under (\ref{susy}) because of the inclusion of the boundary terms $\delta(y){\bar\Psi}^{[\mu} {\Psi}^{\nu]}$. We want to show that with fixed boundaries the partition function of the string theory is the same as the expectation value of products of super-Wilson-loops in Maxwell theory:
\be
\int\left( \prod_{j} {\mathscr{D}{\bf X}_j\over Z_0}\right)\, e^{-S_s}
=\int {{\mathscr{D}}A\over N}\,e^{-S_{gf}}\,\prod_j
W_s[A]\label{sexpWL}
\ee
which is a functional of the boundary data consisting of world-line variables associated with the closed loops. In computing the left hand-side we expand in powers of the contact interaction. Representing the delta-function as a Fourier integral reduces the problem to the expectation value of multiple insertions of 
\be
\int d^2z\,d^2\theta\,V^{\mu\nu}(k),\quad {\rm with}\quad V^{\mu\nu}(k)=\bar D {\bf X}^{[\mu} D{\bf X}^{\nu]}\,e^{ik\cdot {\bf X}}\,.
\ee
So we begin with the simplest case of a single insertion on the $j$-th world-sheet and consider the integral
\be
I_j^{\mu\nu}(k)\equiv \int\mathscr{D}{\bf X}_j\, e^{-S_{spin}}\int d^2zd^{2}\theta\left(\bar D {\bf X}^{[\mu} D{\bf X}^{\nu]}\,-\delta(y)\theta\bar{\theta}{\bar\Psi}^{[\mu} {\Psi}^{\nu]}\right)e^{ik\cdot {\bf X}}
\,.
\ee
Athough classically superconformally invariant the insertion acquires an anomalous dimension that would take it off-shell unless $k$ were null, so conventionally $\delta$-function contact interactions do not appear in critical string theory. However we argue that the same self-contraction of the exponential that gives rise to this also suppresses the insertion for all points $z$ that are not close (on the scale of the short-distance regulator) to the boundary. Because of the  Dirichlet boundary conditions however, points close to the boundary 
make a finite scale independent contribution as we will see. Set ${\bf X}={\bf X}_c+\tilde{\bf X}$ with ${\bf X}_c$ a classical piece satisfying the boundary conditions (\ref{bcon}) and 
Euler-Lagrange equations $\bar D D{\bf X}_c=0$, and $\tilde{\bf X}$
a quantum fluctuation. Integrating over $\tilde{\bf X}$ gives
\bq&
e^{-S_{spin}[{\bf X}_c]-S_L}\int d^2z\Big(\int d^2\theta\,e^{ik\cdot {\bf X}_c-\pi\alpha' k^2 G_0}
\left(\bar D {\bf X}_c^{[\mu} D{\bf X}_c^{\nu]}
-2\pi\alpha'\left(\bar D {\bf X}_c^{[\mu} (DG)_0ik^{\nu]}+(\bar D G)_0ik^{[\mu} D{\bf X}_c^{\nu]}
\right)\right)\nonumber\\
&
-\delta(y)e^{ik\cdot X_{c}}{\bar\Psi}^{[\mu} {\Psi}^{\nu]}\Big)
\label{into}
\eq
where $S_L$ contains the logarithms of functional determinants that give rise to the super-Liouville action. $G$ is the Green function satisfying:
\be
-\bar D DG=(\theta_1-\theta_2)(\bar\theta_1-\bar\theta_2)\delta^2(z_1-z_2),\quad G=0\,\,\, {\rm if}\,\,\,y_1=0\,\,\,{\rm
and}\,\,\,\theta_1=\bar\theta_1\,\,\,{\rm or}\,\,\,y_2=0\,\,\,{\rm
and}\,\,\,\theta_2=\bar\theta_2\label{bc}
\ee
The subscript $0$ on $G$ and its derivatives denotes that  they should be evaluated at coincident points, i.e. $z_1=z_2$, $\theta_1=\theta_2$, $\bar\theta_1=\bar\theta_2$, however this is singular so $G$ must be regulated. We choose a heat-kernel regulator and replace $G$ by
\bq
&G^\epsilon=-f\left(\sqrt{\vphantom{z^{R}}z_{12}\bar z_{12}/ \epsilon}\right)+f\left(\sqrt{z_{12}^R\bar z_{12}^R/ \epsilon}\right)\nonumber\\
&z_{12}=z_1-z_2-\theta_1\theta_2,\,\,\,\bar z_{12}=\bar z_1-\bar z_2-\bar\theta_1\bar\theta_2,\,\,\,
z_{12}^R=z_1-\bar z_2-\theta_1\bar \theta_2,\,\,\,\bar z_{12}^R=\bar z_1-z_2-\bar\theta_1\theta_2,\,.
\eq
with $\epsilon$ a short distance cut-off to be taken to zero at the end of calculations and
\be
f\left(s\right) = \int_{1}^{\infty} \frac{d\tau}{4\pi\tau}
\left(1-\exp{\left(-\frac{s^2}{\tau}\right)} \right),
\ee
so that
\be
-\bar D DG^\epsilon=(\theta_1-\theta_2)(\bar\theta_1-\bar\theta_2){e^{-z_{12}\bar z_{12}/ \epsilon}\over 4\pi\epsilon}
-(\theta_1-\bar\theta_2)(\bar\theta_1-\theta_2){e^{-z_{12}^R\bar z_{12}^R/ \epsilon}\over 4\pi\epsilon}
\ee
For points in $H$ this is a regularisation of Green's equation. $G_\epsilon$ satisfies the boundary conditions (\ref{bc}). Furthermore this regulator is invariant under the residual supersymmetry (\ref{susy}) when we take the scale of the world-sheet metric to be constant, which will be sufficient for our computations. Using this we obtain
\be
G^\epsilon_0=f\left(-{(2iy-\theta\bar\theta)/\sqrt\epsilon}\right),\quad
(DG^\epsilon)_0=(\bar D G^\epsilon)_0={i\over 2}(\theta-\bar\theta){\partial f(2y/\sqrt\epsilon)\over\partial y}
\ee
Expanding the exponential term in (\ref{into}) in powers of $\theta$ gives
\be
e^{-\pi\alpha' k^2 G_0}=\left(1+{i\over 2}\theta\bar\theta{\partial\over\partial y}\right)e^{-\pi\alpha' k^2 f(2y/\sqrt\epsilon)}\,.
\ee
When $s$ is large $f(s)\approx (\log s)/2\pi$ so, for values of $k^2$ that are fixed as the cut-off is removed, this exponential suppresses the integrand in (\ref{into}) at all points in $H$ apart from those that are close (in terms of $\epsilon$) to the boundary. Consider the behaviour at points for which $0<y<\Lambda$ where $\Lambda\downarrow 0$ as $\epsilon\downarrow 0$ but $\Lambda^2/\epsilon$ diverges.  Here we can replace the classical field ${\bf X}_c$, which varies slowly on the scale of $\epsilon$, by its boundary value. Thus the first term in (\ref{into}) is given for small $\epsilon$ as
\be
\int d^2z\,d^2\theta\,e^{ik\cdot {\bf X}_c-\pi\alpha' k^2 G_0}
\bar D {\bf X}_c^{[\mu} D{\bf X}_c^{\nu]}=-2i
\int dx\,d^2\theta\,e^{ik\cdot {\bf X}_c}\bar D {\bf X}_c^{[\mu} D{\bf X}_c^{\nu]}\int_0^\Lambda dy \,\left(1+{i\over 2}\theta\bar\theta{\partial\over\partial y}\right)e^{-\pi\alpha' k^2 f(2y/\sqrt\epsilon)}\label{clas}\,.
\ee
$f(s)$ is monotonically increasing so $\left|\int_0^\Lambda dy \exp (-\pi\alpha' k^2 f(2y/\sqrt\epsilon))\right|<\Lambda$ which goes to zero as the cut-off is removed. Given that $f(0)=0$, it follows that, as $\epsilon\downarrow 0$
\be
\int_0^\Lambda dy \,\left(1+{i\over 2}\theta\bar\theta{\partial\over\partial y}\right)e^{-\pi\alpha' k^2 f(2y/\sqrt\epsilon)}
\rightarrow
-{i\over 2}\theta\bar\theta\,.
\ee
and upon integrating over the anti-commuting variables (\ref{clas}) becomes
\be
\int dx\,e^{ik\cdot { X}_c}\bar\Psi^{[\mu}_c\Psi^{\nu]}_c
\ee
which cancels against the boundary term in (\ref{into}). Similarly the remaining terms in (\ref{into}) are, for small $\epsilon$,
\bq
&&\int d^2z\,d^2\theta\,e^{ik\cdot {\bf X}_c-\pi\alpha' k^2 G_0}
\left(\bar D {\bf X}_c^{[\mu} (DG)_0ik^{\nu]}+(\bar D G)_0ik^{[\mu} D{\bf X}_c^{\nu]}
\right)\nonumber\\
&&
=
-\int dx\,d^2\theta\,e^{ik\cdot {\bf X}_c}\left(\bar D {\bf X}_c^{[\mu} ik^{\nu]}-ik^{[\mu} D{\bf X}_c^{\nu]}
\right)
{(\theta-\bar\theta)\over\pi\alpha' k^2}
\int_0^\Lambda dy \,{\partial \over\partial y}\,e^{-\pi\alpha' k^2 G_0}
\label{intoo}
\eq
The $y$-integral tends to unity as $\epsilon\downarrow 0$ and the integral with respect to $\theta$ leaves 
\be
{1\over\pi\alpha' k^2}\int dx\,e^{ik\cdot { X}_c}\left(ik\cdot\left(\Psi_c+\bar\Psi_c\right)(\Psi_c+\bar\Psi_c)^{[\mu}+{\partial X_c^{[\mu}/\partial x}\right)ik^{\nu]}
\ee
so, using (\ref{bcon}) we obtain the $\epsilon\downarrow 0$ limit of (\ref{into}) as
\be
I_j^{\mu\nu}=-2e^{-S_{spin}[{\bf X}_c]-S_L}\int dx\,e^{ik\cdot w}
\left({d w_c^{[\mu}/d x}+\sqrt h \,ik\cdot\psi\psi^{[\mu}\right)ik^{\nu]}/k^2
\label{result}
\ee
Now in this expression the length scale $\sqrt{\alpha'}$ appears only in $S_{spin}[{\bf X}_c]$ so we can remove this classical action by taking the tensionless limit $l/\sqrt{\alpha'}\rightarrow 0$, where $l$ is a measure of the size of the closed loop $B$. Additionally we can remove $S_L$ by assuming that there are sufficient additional internal degrees of freedom. $S_L$ contains the super-Liouville degrees of freedom, i.e. the scale of the metric and its super-partner on the world-sheet. These degrees of freedom have not appeared in our result for $I_j^{\mu\nu}$, even though we have not restricted $k^2$ by a mass-shell condition. We have effectively worked with a constant world-sheet metric and absorbed the scale into the cut-off $\epsilon$. The finiteness of $I$ as the cut-off is removed demonstrates that $I_j^{\mu\nu}$ is independent of this constant 
scale. Spatial variations of the scale on the world-sheet would only contribute at higher order in $\epsilon$ and so vanish as this cut-off is removed therefore $I_j^{\mu\nu}$ is independent of this scale. So even if there are no additional degrees of freedom to cancel $S_L$ the super-Liouville theory decouples (assuming that the world-line and world-sheet metrics are treated as independent of each other).

Using (\ref{result}) we can evaluate the effect of the interaction to leading order when we average over distinct world-sheets:
\bq
&&\int{\mathscr{D}{\bf X}_j\over Z_0}{\mathscr{D}{\bf X}_{j'}\over Z_0}
\, e^{-S_{spin}[{\bf X}_j]-S_{spin}[{\bf X}_{j'}]}\,
S_{int}[{\bf X}_j,{\bf X}_{j'}]
=q^2\int {d^dk\over (2\pi)^d} \,I_j^{\mu\nu}(k)
I_{j'}^{\mu\nu}(-k)\nonumber\\
&&
=q^2\int {d^dk\over (2\pi)^d}
\int dx\,dx'\,{e^{ik\cdot(w-w')}\over k^2}\left({dw\over dx}+\sqrt h\, 
\psi\cdot ik\, \psi\right)\cdot\left({dw'\over dx'}+\sqrt {h'} 
\,\psi'\cdot ik\, \psi'\right)\label{reso}
\eq
which we recognise as the order $q^2$ contribution to the expectation value of two super-Wilson loops in QED. This verifies (\ref{sexpWL}) to leading order when distinct world-sheets are involved. We now argue that this extends to all orders. This will rely on our procedure (namely the action, interaction and regulator) preserving the residual supersymmetry (\ref{susy}). A general term in the expansion of 
(\ref{sexpWL}) will involve multiple insertions at various points $z_r$ on each world-sheet so we need to compute
\be
I^{\mu_1\nu_1..}(k_{1},..)\equiv \int\mathscr{D}{\bf X}\, e^{-S_{spin}}\int d^2z_1d^{2}\theta_{1}..\prod_r\left.\left(\bar D {\bf X}^{[\mu_r} D{\bf X}^{\nu_r]}-\delta(y_r)\theta_{r}\bar{\theta}_{r}{\bar\Psi}^{[\mu_r} {\Psi}^{\nu_r]}\right)e^{ik_r\cdot {\bf X}}\right|_{z_r}
\,.
\ee
When all the points $z_r$ are separated by more than $\Lambda$
the computation parallels that for a single insertion. The exponential factors $\exp(-\pi\alpha'k^2G_0)$ that appear after integrating over ${\bf X}_j$ suppress the contribution of insertions except when $y_r<\Lambda$ and points close to the boundary result in a product of terms like (\ref{result}). These terms then yield the required result (\ref{sexpWL}). However, when some of the insertions approach each other divergences might arise that would spoil the above argument. We will show that the residual supersymmetry prevents this.

Consider a set of $n+1$ insertions all being within $\Lambda$ of each other, but separated by more than $\Lambda$ from the others. Using Wick's theorem, their contribution may be replaced by a sum of terms involving contractions between the set and normal ordered terms (denoted by colons) which are yet to be contracted with operators outside the set. E.g. for two insertions
\bq
&&
\bar D {\bf X}^{[\mu_1} D{\bf X}^{\nu_1]}\,e^{ik_1\cdot {\bf X}}\Big|_{z_1}\,\bar D {\bf X}^{[\mu_2} D{\bf X}^{\nu_2]}\,e^{ik_2\cdot {\bf X}}\Big|_{z_2}\nonumber\\
&&
=\,:\bar D {\bf X}^{[\mu_1} D{\bf X}^{\nu_1]}\Big|_{z_1}\,\bar D {\bf X}^{[\mu_2} D{\bf X}^{\nu_2]}\Big|_{z_2}\,e^{ik_1\cdot {\bf X}(z_1)+ik_2\cdot {\bf X}(z_2)}:e^{-\pi\alpha'\sum k_r\cdot k_s G^{\epsilon}(z_r,z_s)}\nonumber\\
&&
+\,:\bar D {\bf X}^{[\mu_1} D{\bf X}^{\nu_1]}\Big|_{z_1}\,\bar D {\bf X}^{[\mu_2} ik_2^{\nu_2]}\Big|_{z_2}\,D_2G^\epsilon(z_2,z_1)e^{ik_1\cdot {\bf X}(z_1)+ik_2\cdot {\bf X}(z_2)}:e^{-\pi\alpha'\sum k_r\cdot k_s G^{\epsilon}(z_r,z_s)}\nonumber\\
&&+...\nonumber\\
&&
-\delta^{\mu_1\nu_1}_{\mu_2\nu_2}\left(D_1D_2G^{\epsilon}\,\bar D_1\bar D_2G^{\epsilon}+D_1\bar D_2G^{\epsilon}\,\bar D_1 D_2 G^{\epsilon}\right):e^{ik_1\cdot {\bf X}(z_1)+ik_2\cdot {\bf X}(z_2)}:e^{-\pi\alpha'\sum k_r\cdot k_s G^{\epsilon}(z_r,z_s)}
\label{twopoint}
\eq
Furthermore the terms inside the colons can be expanded around the position of, say, the first insertion, so, in the general case
\bq
\prod_{r=1}^{n+1}\left(\left.\bar D {\bf X}^{[\mu_r} D{\bf X}^{\nu_r]}\,e^{ik_r\cdot {\bf X}}\right|_{z_r}\right)
&=&:\left.\left(\prod_{r=1}^{n+1}\bar D {\bf X}^{[\mu_r} D{\bf X}^{\nu_r]}\right)\right|_{z_1}\,e^{i(\sum k_r)\cdot {\bf X}(z_1)}:e^{-\pi\alpha'\sum k_r\cdot k_s G^{\epsilon}(z_r,z_s)}
\nonumber\\
&&+\,...\nonumber\\
&&+F^{\mu_1..\nu_{n+1}}(z_1,..,z_{n+1}):e^{i(\sum k_r)\cdot {\bf X}(z_1)}:e^{-\pi\alpha'\sum k_r\cdot k_s G^{\epsilon}(z_r,z_s)}
\label{ope}
\eq
Now
\bq
G^{\epsilon}(z_r,z_s)&&=-f\left(\sqrt{\vphantom{Z}z_{rs}\bar z_{rs}/\epsilon}\right)+
f\left(\sqrt{z_{rs}^R\bar z_{rs}^R/\epsilon}\right)\nonumber\\
&&=
-f\left(\sqrt{\vphantom{Z}z_{rs}\bar z_{rs}/\epsilon}\right)+
{1\over4\pi}\log\left({(2iy_1-\theta_r\bar\theta_s)(-2iy_1-\bar\theta_r\theta_s)\over\epsilon}\right)+O(\Lambda/y_{1})
\label{Gsusy}
\eq
The most divergent terms in (\ref{twopoint}) and (\ref{ope}) are contained in the coefficient $F$ which consists of $2(n+1)$ derivatives, $D$ and $\bar D$, acting on various combinations of $G^{\epsilon}(z_r,z_s)$. The leading terms are those in which the derivatives all act on the $f(\sqrt{z_{rs}\bar z_{rs}/\epsilon})$ parts. To see this scale all the relative co-ordinates $z_r-z_s$ (but not $z_1$ or $\bar z_1$) and the $\theta_r$, $\bar\theta_r$:
\be
z_r-z_s\rightarrow \sqrt \epsilon (z_r-z_s),\quad\theta_r\rightarrow \epsilon^{1/4} \theta_r,\quad\bar\theta_r\rightarrow \epsilon^{1/4} 
\bar\theta_r, \quad {\rm so}\quad
f(\sqrt{z_{rs}\bar z_{rs}/\epsilon})\rightarrow f(\sqrt{z_{rs}\bar z_{rs}})
\ee
and
\be
\quad D\rightarrow \epsilon^{-1/4} D,\quad \bar D\rightarrow \epsilon^{-1/4} \bar D, 
\quad d^2 z_r d^2\theta_r\rightarrow \epsilon^{1/2}d^2 z_r d^2\theta_r \quad  r>1,\quad d^2 z_1 d^2\theta_1\rightarrow \epsilon^{-1/2}d^2 z_1 d^2\theta_1 
\ee
so the integral with respect to $\prod_{r} d^2 z_r d^2\theta_r $ of the term containing $2(n+1)$ derivatives, $D$ and $\bar D$, acting on $f(\sqrt{z_{rs}\bar z_{rs}/\epsilon})$  scales into 
$1/\epsilon$ multiplied by an integral independent of $\epsilon$. This depends on the $k_r$ in a potentially complicated way but the ${\bf X}$ dependence is quite simple so, after the integral over the relative co-ordinates and the $\theta_r$,$\bar\theta_r$ are done we are left with
\be
{1\over\epsilon}\tilde F^{\mu_1..\nu_{n+1}}(k_1,..,k_{n+1})\,
\int d^2z_1:e^{iK\cdot { X}(z_1)}:\left({\epsilon\over y_{1}^2}\right)^{\alpha'K^2/4}\,,
\ee
where
\be
\tilde F^{\mu_1..\nu_{n+1}}(k_1,..,k_{n+1})=
\int d^2\theta_1\left(\prod_{j=2}^{n+1} d^2z_j\,d^2\theta_j\right)\,F^{\mu_1..\nu_{n+1}}(z_1,..,z_{n+1})
e^{\pi\alpha'\sum k_r\cdot k_s f(\sqrt {z_{rs}\bar z_{rs}})}
\,,\quad K=\sum_{j=1}^{n+1} k_j\,.
\ee
This is not invariant under the residual supersymmetry and so must vanish. There can be no subleading terms of order $\epsilon^{-3/4}$ since their super-field content would have to be fermionic to generate the factor of $\epsilon^{1/4}$ needed. The next non-trivial terms are of order $1/\sqrt\epsilon$ and using rotational symmetry the only possibility is an $X$-dependence proportional to
\be
{c^{\rho\sigma}\over\sqrt\epsilon}\int d^2z_1:\bar\Psi^\rho\Psi^\sigma e^{iK\cdot { X}(z_1)}:\left({\epsilon\over y_{1}^{2}}\right)^{\alpha'K^2/4}\,,
\label{div}
\ee
This too changes under the residual supersymmetry, although if $c^{\rho\sigma}=K^\rho K^\sigma$ its variation is
proportional to the variation of the boundary term $\epsilon^{-1/2}\int dx \,\exp(ik\cdot w)$, so if this boundary term were also generated as the insertions approached each other close to the boundary then there would be the possibility of a divergence. However a term like (\ref{div}) does not appear because $k\cdot \bar\Psi \,k\cdot \bar\Psi$ can only be generated by expanding the $\theta\bar\theta$ terms in the exponent so the coefficient of the term would be
\be
\int d^2\theta_1\left(\prod_{j=2}^{n+1} d^2z_j\,d^2\theta_j\right)\,F^{\mu_1..\nu_{n+1}}(z_1,..,z_{n+1})e^{\pi\alpha'\sum k_r\cdot k_s f(\sqrt {z_{rs}\bar z_{rs}})}\,\bar\theta_r\theta_s \label{poss}
\ee
(with the result being independent of the choice of $r,s$).
By counting $\theta$s we can see that this vanishes: there are $n+1$ derivatives $D$ and $n+1$ derivatives $\bar D$ acting on $f$ contained in $F$. Each of these produces terms with the same number of $\theta$ and $\bar\theta$ counting mod 2, resulting in $n+1$ (mod 2) $\theta$ and $n+1$ (mod 2) $\bar\theta$, not the $n$ $\theta$  and $n$ $\bar\theta$ needed for (\ref{poss}) to be non-zero.

The only other divergence we could encounter is at order $\epsilon^{1/4}$ but the field content of these terms would also be fermionic and so cannot be present. Since $K$ is real the remaining contributions in the expansion (\ref{ope}) are suppressed by the common factor $\left(\epsilon / y_{1}^{2}\right)^{\alpha' K^{2}/4}$ arising from the second term in (\ref{Gsusy}) 
which vanishes as the regulator is removed for all $K^{2}$ except those close to zero (in terms of $\epsilon$). Since $K$ is ultimately to be integrated over we also need to consider the contribution of these small values, however for $\alpha'$ large and $\epsilon$ small this factor behaves as
\begin{equation}
	\frac{\delta\left(K^{2}\right)}{\left(\frac{1}{2} \alpha^{\prime} \ln{\frac{y_{1}}{\epsilon}}\right)^{\frac{D}{2}}}
\end{equation}
and so is also suppressed in the tensionless limit -- see \cite{us} for further detail. We conclude that no divergent terms can be generated by insertions that approach each other far from the boundary. 

As the insertions approach each other close to the boundary the second term in $G^\epsilon$ varies rapidly so we have to consider its variation too by scaling $y_1$ in addition to the other variables. Consequently in the integral of (\ref{ope}) there are potential terms of order $1/\sqrt \epsilon$, but these take the form $\epsilon^{-1/2}\int dx \,\exp(iK\cdot w)$ which we have already dealt with. We can ignore the $\mathcal{O}\left(\epsilon^{-1/4}\right)$ contribution since it would have fermionic super-field content so the next order in $\epsilon$ consists of finite terms. There is one candidate that is invariant under the residual supersymmetry and so could potentially occur, and that is the electromagnetic coupling:
\be
\int dx~ e^{iK\cdot w} \left( dw^\mu/dx+iK\cdot (\Psi+\bar\Psi)
(\Psi+\bar\Psi)^\mu\right)\,\label{posso}
\ee
Potentially this could arise from one of the $D{\bf X}$, say the $q$-th, being replaced by their classical value $D{\bf X}_c^{\mu_q}$ which would generate the $dw^\mu/dx$ piece, so $\mu=\mu_q$. However if we contract the integral of an insertion with $k$ the result is a boundary term that does not contain the quantum variables:
\be
k^\mu\int \,d^2 zd^2\theta\left(\bar D {\bf X}^{[\mu} D{\bf X}^{\nu]}-\delta(y)\theta\bar{\theta}{\bar\Psi}^{[\mu} {\Psi}^{\nu]}\right)e^{ik\cdot {\bf X}}=\int dx ~e^{ik\cdot X} \left({dX^\nu\over dx} + ik\cdot\left(\Psi + \bar{\Psi}\right) \left(\Psi + \bar{\Psi}\right)^{\nu}\right)
\label{gausso}
\ee
which factors out of the sum of normal ordered terms due to the other insertions in the set. So this boundary integral of the $q$-th field would have to factor out of the contraction of (\ref{posso}) with $k_{q}$ which is not possible because it contains only one field integrated around the boundary\footnote{(\ref{gausso}) is related to Gauss' law.}.
In conclusion, supersymmetry prevents divergences appearing when the insertions approach each other, consequently (\ref{reso}) exponentiates, leading to (\ref{sexpWL}). As a final step we integrate over the world-sheet metric and boundaries weighted by the world-line action 
\be
\int\left( \prod_{j}^n {\mathscr{D}(g,{\bf X},w,\psi,h,\chi)_j\over Z_0}\right)\, e^{-S_s-S_{BdVH}}
=\int \left(\prod_{j}^n {\mathscr{D}(w,\psi,h,\chi)_j}\right){{\mathscr{D}}A\over N}\,e^{-S_{gf}-S_{BdVH}}\,\prod_j
W_s[A].\label{sexpWL2}
\ee
On summing over $n$ this expresses the equality of the partition functions of QED and of tensionless spinning strings
with contact interactions. Following Strassler we can include a background gauge field on the world-lines to source photon amplitudes, and, as explained in  (\cite{us}) the Green functions for the charged particles are obtained by including open world-lines with appropriate boundary conditions at their ends.
\section{Concluding remarks}
We have shown that QED can be described by the tensionless limit of spinning strings with contact interactions. World-sheet supersymmetry has underpinned the consistency of the construction and indicates that the model has a preference for spinor matter. The string world-sheets are the trajectories of lines of electric flux connected to electric charges at their ends, a picture reminiscent of the old dual resonance model.  Integrating over these produces the electromagnetic super-Wilson loops (associated with world-sheet boundaries) necessary to describe the electromagnetic coupling of spinor matter. These spinning strings are physically very different from the fundamental strings of quantum gravity. They interact via $\delta$-functions on the world-sheet which are not present in critical string theory because they naively break super-conformal invariance but they contribute here because of the different boundary conditions. Furthermore, because the string length scale is taken large in comparison to the 
size of the Wilson loops the strings themselves can be very large, possibly macroscopic. Potentially the model might be distinguished observationally from conventional QED where the electromagnetic degrees of freedom are photons by direct observation of the strings themselves, or by string-like corrections to QED that might set in at large distances. QED is, of course, an effective theory that emerges from the spontaneous breaking of the gauge symmetry of the standard model, and so the string model described here is similarly an effective theory. Understanding how it relates to the more fundamental Weinberg-Salam theory will require further work to obtain the generalisation to non-Abelian gauge theory.

Both authors are grateful  to STFC: PM for support under the Consolidated Grant ST/J000426/1, and JPE for a studentship. This research is also supported in part by the the Marie Curie network GATIS 
(gatis.desy.eu) of the European Union's Seventh Framework Programme FP7/2007-2013/ under REA Grant Agreement No 317089.




\begin{thebibliography}{00}


\bibitem{Faraday}
M.~Faraday,
``Thoughts on Ray-vibrations,'' 
Philosophical Magazine, 1846, vol. xxviii, p. 345
reprinted in ``Experimental Researches in Chemistry and Physics,
ISBN 0-85066-841-7.

\bibitem{Dirac}
P. ~A.~M.~Dirac,
``Gauge-Invariant Formulation of Quantum Electrodynamics,''
Canadian Journal of Physics, 1955 {\bf 33} 650.


\bibitem{Woolley}
R.~G.~ Woolley
Molecular Physics, {\bf 22} (1971) 1013,~
Proc. Roy. Soc. London {\bf A321} (1971) 557. 

\bibitem{NO}
	H.~B.~Nielsen and P.~Olsen,
	``Local field theory of the dual string,''
	Nucl.\ Phs.\ B {\bf 57} (1973) 367.
	
\bibitem{diracM}
	P.~A.~M.~Dirac,
	''The theory of magnetic poles,''
	Phys.\ Rev.\ {\bf{74}} (1948) 817.
	
	\bibitem{Nambu}
	Y.~Nambu,
	``Strings, monopoles, and gauge fields,''
	Phys.\ Rev.\ {\bf D10} (1974) 4262.

	\bibitem{Baker}	
	M.~Baker and R.~Steinke
	``Effective String Theory of Vortices and Regge Trajectories,''
	[arXiv:hep-ph/0006069]
	
	\bibitem{kalbR}
	M.~Kalb and P.~Ramond,
	``Classical direct interstring action,''
	Phys\. Rev.\ {\bf D9} (1974) 2273.
	
	\bibitem{NRH}
	R.~I.~Nepomechie, M.~A.~Rubin and U.~Hosotani
	``A new formulation of the string action,''
	Phys.\ Lett.\ {\bf B105} (1981) 457.
\bibitem{Mansfield:2011eq}
  P.~Mansfield,
``Faraday's Lines of Force as Strings: from Gauss' Law to the
Arrow of Time,''
  JHEP {\bf 1210} (2012) 149
  [arXiv:hep-ph/1108.5094].

\bibitem{Polyakov}
  A.~M.~Polyakov,
  ``Quantum geometry of bosonic strings,''
  Phys.\ Lett.\   {\bf B103} (1981) 207.


\bibitem{Strassler:1992zr}
  M.~J.~Strassler,
``Field theory without Feynman diagrams: One loop effective
actions,''
  Nucl.\ Phys.\  B {\bf 385} (1992) 145
  [arXiv:hep-ph/9205.205].

\bibitem{us}
J.~P.~Edwards, and  P.~Mansfield
``Delta-function Interactions for the Bosonic and Spining Strings and the Generation of Abelian Gauge Theory."
In preparation

\bibitem{Schubert:2001he}
  C.~Schubert,
  ``Perturbative quantum field theory in the string inspired formalism,''
  Phys.\ Rept.\  {\bf 355} (2001) 73
  [hep-th/0101036].

\bibitem{Ilderton:2014mla}
  A.~Ilderton,
  ``Localisation in worldline pair production and lightfront zero-modes,''
  arXiv:hep-th/1406.1513].

\bibitem{Brink:1976uf}
  L.~Brink, P.~Di Vecchia and P.~S.~Howe,
``A Lagrangian Formulation Of The Classical And Quantum Dynamics
Of Spinning
  Particles,''
  Nucl.\ Phys.\  B {\bf 118} (1977) 76.


\bibitem{Friedan:1985ge}
  D.~Friedan, E.~J.~Martinec and S.~H.~Shenker,
  ``Conformal Invariance, Supersymmetry and String Theory,''
  Nucl.\ Phys.\ B {\bf 271} (1986) 93.


\end{thebibliography}


\end{document}